\documentclass{appolb}
\usepackage{epsfig,graphicx}

\begin{document}

\pagestyle{plain}

\title{Cross-correlations in Warsaw Stock Exchange}

\author{R.Rak$^a$, J.~Kwapie\'n$^b$, S.~Dro\.zd\.z$^{a,b}$, 
P.~O\'swi\c ecimka$^b$
\address{$^a$Institute of Physics, University of Rzesz\'ow, PL--35-310
Rzesz\'ow, Poland\\
$^b$Institute of Nuclear Physics, Polish Academy of Sciences, \\
PL--31-342 Krak\'ow, Poland}}

\maketitle

\begin{abstract}

We study the inter-stock correlations for the largest companies listed on
Warsaw Stock Exchange and included in the WIG20 index. Our results from
the correlation matrix analysis indicate that the Polish stock market can
be well described by a one factor model. We also show that the stock-stock
correlations tend to increase with the time scale of returns and they
approach a saturation level for the time scales of at least 200 min, i.e.
an order of magnitude longer than in the case of some developed markets.
We also show that the strength of correlations among the stocks crucially
depends on their capitalization. These results combined with our earlier
findings together suggest that now the Polish stock market situates itself
somewhere between an emerging market phase and a mature market phase.

\end{abstract}


\section{Introduction}

Since the pioneering work of Markowitz in 1950s~\cite{markowitz52}, the
financial cross-correlations are constantly a subject of extensive studies
both at the theoretical and practical levels due to their fundamental
relation to risk management and portfolio investing. In the field of
econophysics, an interest in this type of correlations arose after it had
been shown that they can be described~\cite{laloux99,plerou99,plerou02} in
the framework of Random Matrix Theory (RMT)~\cite{mehta}, expressing both
a kind of universality and significant deviations from it. Stock market
cross-correlations are typically quantified in terms of a correlation
matrix, created for a set of $N$ time series representing returns of
different stocks. From this point of view, evolution of a stock market can
be decomposed into $N$ independent modes associated with eigenvalues of
the correlation matrix. It occurs that a vast majority of these
eigenvalues are concordant with the eigenvalue distribution of the
relevant random matrix ensemble (the Wishart
ensemble)~\cite{laloux99,sengupta99}, what according to a common belief
suggests that these RMT modes do not carry any market-specific information
beyond being a pure noise. Validity of this belief, however, has recently
been challenged in some works~\cite{kwapien06}. As regards the remaining
minority of the eigenvalues which deviate from the RMT predictions, there
is a general agreement that they express the actual non-random linear
dependencies between the price fluctuations of different assets. Their
particular number depends on a market and the number of analyzed stocks,
but in each case there is an eigenvalue that strongly dominates,
developing an "energy gap" that separates it from the subsequent
eigenvalues. This peculiar eigenvalue is associated with a strongly
delocalized eigenvector and is related to a "market mode", i.e. a
collective evolution of large group of stocks that usually closely mimics
evolution of the market's global index. From this perspective, magnitude
of the largest eigenvalue reflects how collective is the evolution of an
analyzed market. If, apart from the largest eigenvalue, there are also
other eigenvalues which do not agree with the RMT spectrum, they
correspond to smaller groups of interrelated stocks that can be usually
identified with market sectors~\cite{plerou02}. It has been found that the
number of the non-random eigenvalues is highest for the largest, mature
markets like New York, London, Frankfurt etc., while the less capitalized
markets, e.g. the emerging ones, develop the spectrum which consists of a
strongly repelled eigenvalue and the bulk with only minor disagreements in
respect to the RMT prediction~\cite{pan07}. In fact, on small markets
sectors and individual companies are too weak to be considered an optimal
reference for the investors. Instead, the investors trade according to the
behaviour of the whole market or even they blindly follow the moves of
the world's largest markets.

Here we analyze high-frequency data from the Warsaw Stock Exchange and
inspect the correlation matrix eigenspectra calculated for a few selected
groups of stocks. We address the question whether the correlation
properties of the Warsaw stock market still situate it among the emerging
markets or, conversely, it has already matured enough to be considered a
developed market. An inspiration for rising this question is the fact that
the WSE evolution shares some properties (like the broad multifractal
spectra and the returns distributions which can be fitted by the
$q$-Gaussians) with the well-established markets, as our earlier studies
showed~\cite{oswiecimka06,rak07,drozdz07a}.

\section{Methodology}

Our tick-by-tick data covered the period from 17 November 2000 to 30 June
2005 and consisted of 39 stocks that were, at least temporarily, included
during this period in WIG20 index. WIG20 is a capitalization-weighted
index comprising the 20 largest companies traded on WSE.
\begin{figure}
\hspace{1.5cm}
\epsfxsize 11cm
\epsffile{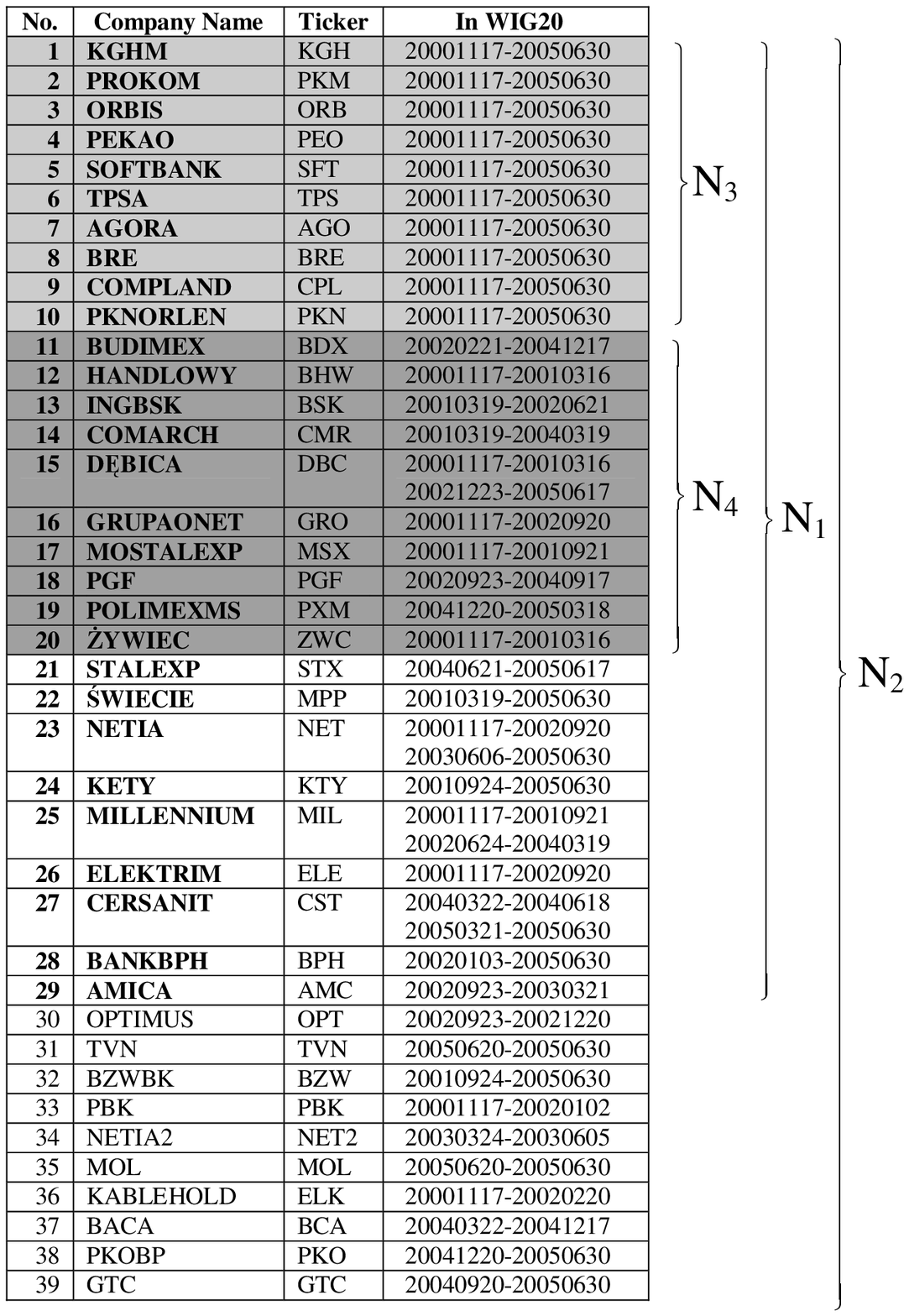}
\caption{Stocks included in WIG20 during the period 17.11.2000 $-$
30.06.2005, divided into four groups $N_1, N_2, N_3, N_4$. See text for
details.}
\end{figure}
Its composition changes from time to time in order to reflect the current capitalization
ranking of the WSE stocks. During the forementioned time interval only 10
stocks were constantly included in WIG20, while each of the remaining 29
stocks was contributing to WIG20 for a shorter period due to falling in
the capitalization ranking or being delisted from WSE.

Although it is associated with a relatively small number of companies,
WIG20 is considered the most important and influential index on WSE. This
is because the companies included in WIG20 are viewed as belonging to a
core of Polish economy. Thus it is not unreasonable to expect that their
statistical and correlation properties differ, at least to some extent,
from the properties of other companies that are not a part of this index.
Also a particular stock may change its behaviour after being included in
or removed from WIG20. In order to investigate this issue we divided our
set of signals into 4 partially overlapping groups (see Figure 1):

(1) Group 1 of $N_1=29$ signals associated with the stocks listed on WSE
over the entire period.

(2) Group 2 of $N_2=20$ signals constituting WIG20 with changing stock
content according to the actual WIG20 basket composition; signals (at the
level of normalized returns) representing the replaced stocks were cut and
joined with the ones representing the replacing stocks.

(3) Group 3 of $N_3=10$ signals for the stocks that were permanently
inluded in WIG20; these are also the companies with the largest
capitalization.

(4) Group 4 of $N_4=10$ signals representing the stocks with the least
capitalization among the ones that were temporarily included in WIG20.

We performed our calculations for each of the above groups independently.

For each individual company $\alpha, \ \alpha = 1,...,N_k$, from the raw
tick-by-tick data we extracted a time series of price evolution
$p_{\alpha}(t_i), \ i=1,...,T$ sampled with 1 min frequency and calculated
the corresponding returns according to the usual definition:
$G_{\alpha}(t_i) = \ln p_{\alpha}(t_i+\tau) - \ln p_{\alpha}(t_i)$, where
$\tau$ is the time lag. After normalizing the time series of returns to
have unit variance and zero mean, the resulting length of each signal was
equal to $T=415,000$.

From $N_k$ time series we construct an $N_k \times T$ data matrix ${\bf
M}$ and the correlation matrix ${\bf C}$ that are related by
\begin{equation}
{\bf C} = (1/T){\bf M} {\bf M}^{\rm T},
\end{equation}
By diagonalizing the correlation matrix
\begin{equation}
{\bf C} {\bf v}^j = \lambda_j {\bf v}^j,
\end{equation}
one obtains a set of its eigenvalues $\lambda_j, \ j=1,...,N_k$ and
eigenvectors ${\bf v}^j = \{ v^j_{\alpha} \}$.

An ensemble of random matrices which can be used as a null hypothesis in
our context is the enseble of Wishart matrices. It offers an analytic
expression for a distribution of eigenvalues, known as the Marchenko-Pastur
formula~\cite{marchenko67,sengupta99}. Here we are interested in its upper
$\lambda_{max}$ and lower $\lambda_{min}$ bounds only:
\begin{equation}
\lambda^{\rm max}_{\rm min} = \sigma^2 (1 + 1/Q \pm 2 \sqrt{1/Q}),
\label{bounds}
\end{equation}
with $Q=T/N_k \ge 1$ and time series variance $\sigma^2=1$.

\section{Results}

Figure 2 shows the eigenvalue spectra for each of the 4 considered groups
of stocks; two time lags are used: $\tau=10$ min (top) and $\tau=360$ min,
i.e. 1 trading day (bottom).
In accordance with the remarks done in the
introductory section, for all the groups and for both time lags the
largest eigenvalue $\lambda_1$ is repelled from the RMT range defined by
Eq.~(\ref{bounds}). Clearly, for Groups 1,2,3 the corresponding shift is
stronger than for Group 4. As regards the rest of the eigenvalues, they
are close to the random matrix region; the observed discrepancies between
their position and the RMT bounds can be attributed to the "squeezing"
effect of large $\lambda_1$ exerted on the smaller
eigenvalues~\cite{laloux99,kwapien06} which are shifted towards zero. The
larger magnitude of $\lambda_1$, the stronger is this effect. Agreement
between the empirical eigenspectrum and the RMT prediction can be
significantly improved by removing a mode associated with $\lambda_1$ from
the analyzed signals (see e.g. ref.~\cite{kwapien06,drozdz07b}).

The largest eigenvalue is related to the temporal evolution of the market
mode. As we see in Figure 2, practically no other eigenvalue considerably
exceeds the RMT upper bound $\lambda_{\rm max}$, what indicates that -
besides the idiosyncratic fluctuations - the market mode is the principal
and unique factor responsible for the behaviour of individual stocks. A
lack of other deviating eigenvalues is the evidence of weakness of a
sectorization in the WSE market.
\begin{figure}
\epsfxsize 15cm
\hspace{-1cm}
\epsffile{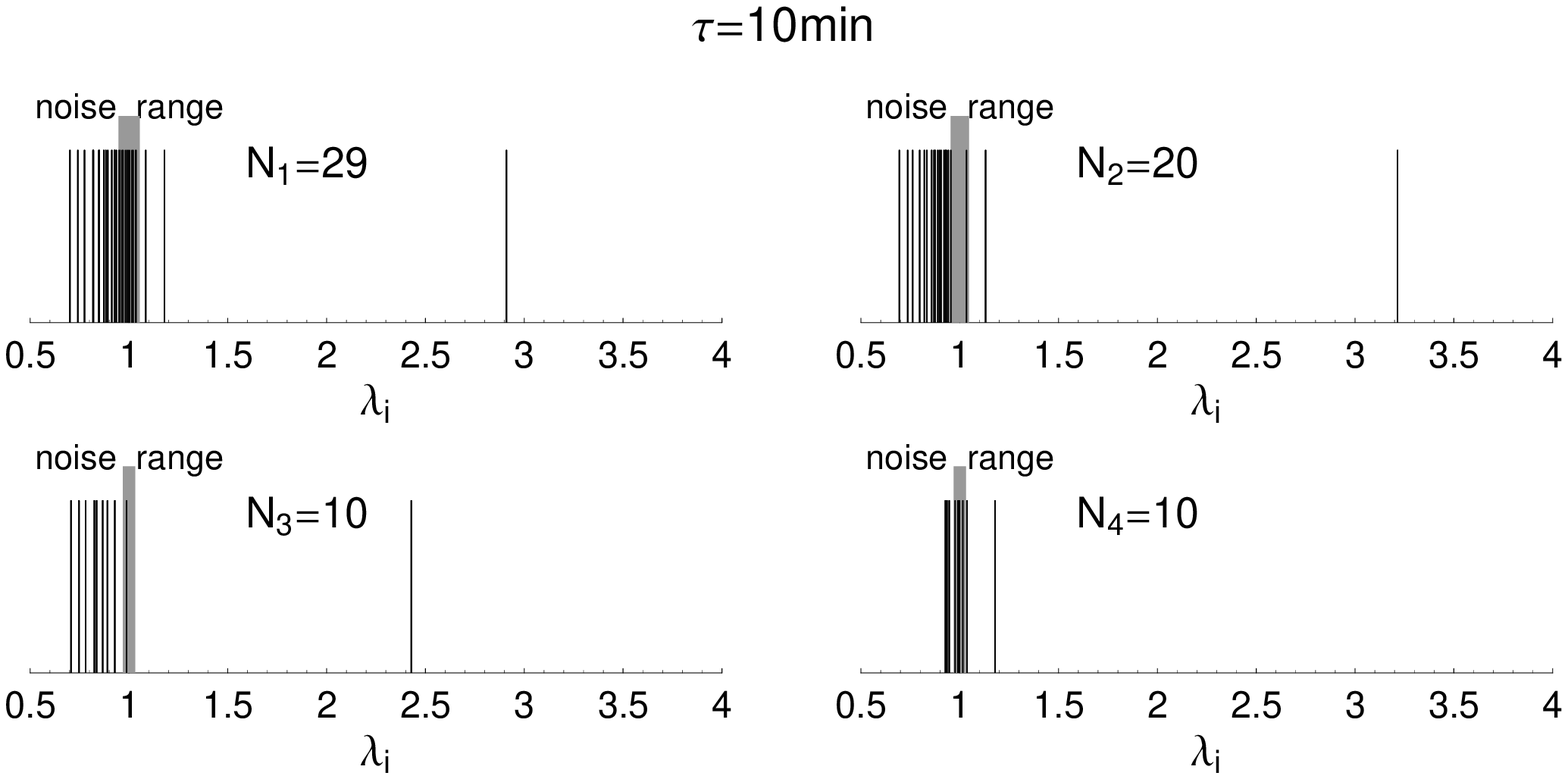}
\vspace{0.5cm}
\epsfxsize 15cm
\hspace{-1.2cm}
\epsffile{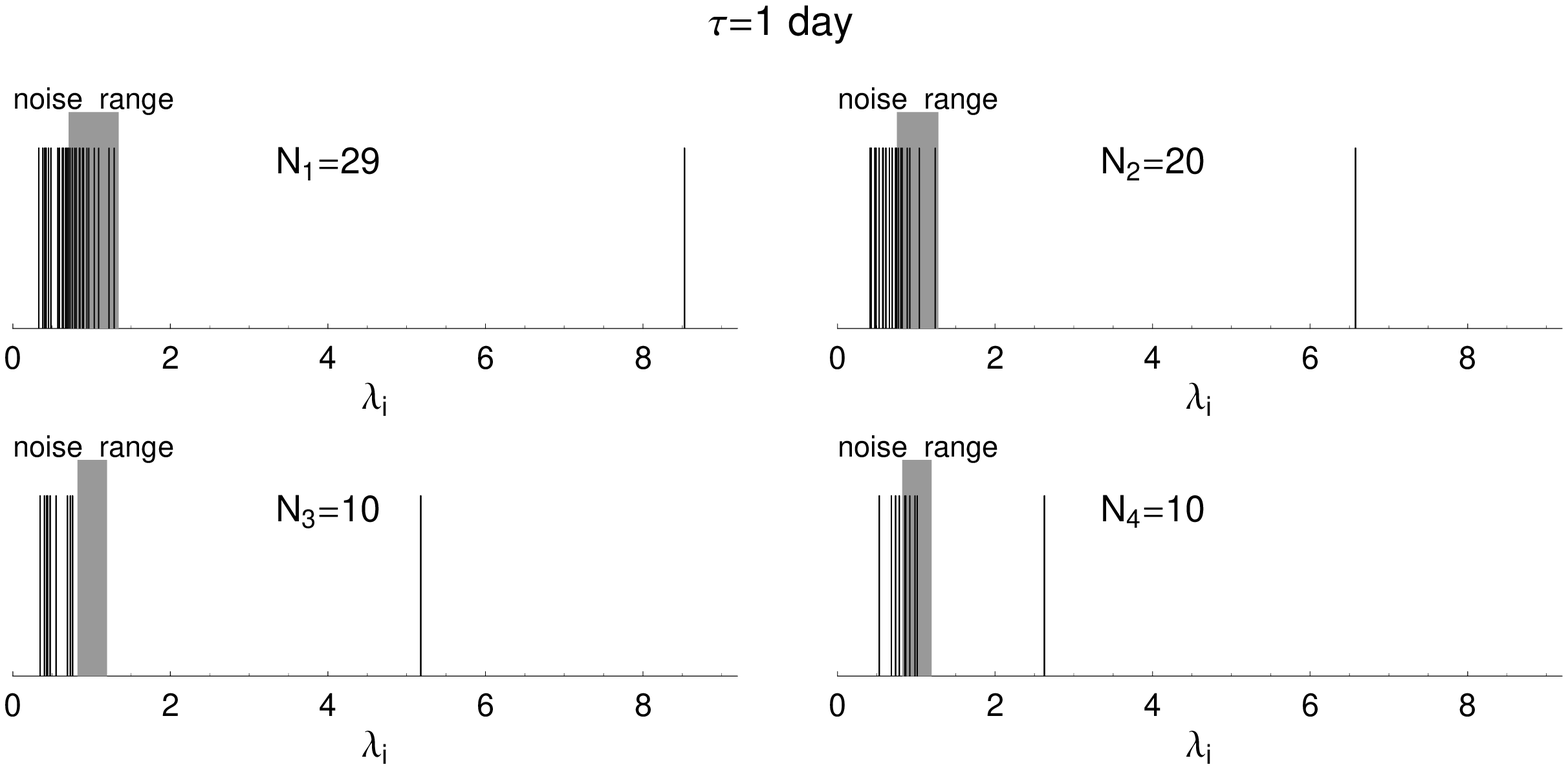}
\caption{Empirical eigenvalue spectrum of the correlation matrix ${\bf C}$
(vertical lines), calculated for 4 group of companies over the period
17.11.2000-30.06.2005. Two time scales are shown: $\tau=10min$ (top) and
$\tau=360min$ (bottom). According to RMT predictions, eigenvalues of a
Wishart matrix have to lie only within the shaded region.}
\end{figure}

The magnitudes of the largest eigenvalues for different groups of stocks
cannot be directly compared based on Figure 2. This is because of a lack
of a proper normalization: the correlation matrices for different groups
have different size and different trace. Thus, in each case we divide
$\lambda_1$ by the matrix trace. As a result we obtain normalized values
of $\lambda_1$ expressing a fraction of the maximum possible magnitude
(equal to the matrix trace) that is absorbed by the largest eigenvalue. It
thus describes the "rigidity" of a given group's temporal evolution. The
corresponding results are exhibited in Figure 3. For both time scales, the
largest normalized magnitude of $\lambda_1$ is observed for Group 3,
comprising the largest companies permanently listed in WIG20. The
difference between this and other groups is especially substantial for the
larger time scale of 1 trading day (Figure 3, bottom panel). This
indicates that the stocks for the largest companies are particularly
strongly coupled with each other. Evidently smaller strength of collective
movements can be seen for Group 2 (representing the actual content of
WIG20), which is associated with the second-largest magnitude of
$\lambda_1$. The stocks within the remaining Groups 3 and 4 are relatively
weakly correlated. In contrast to Groups 1 and 2, which at each moment
comprise the stocks belonging to the WIG20 basket, the stocks from Groups
3 and 4 are not necessarilly included in WIG20 during the whole studied
interval of time: at each particular moment some of them belong to WIG20
and some of them do not. This also means that their capitalization is,
on average, smaller than the one for the stocks from Groups 1 and 2.

\begin{figure}
\epsfxsize 10cm
\hspace{1cm}
\epsffile{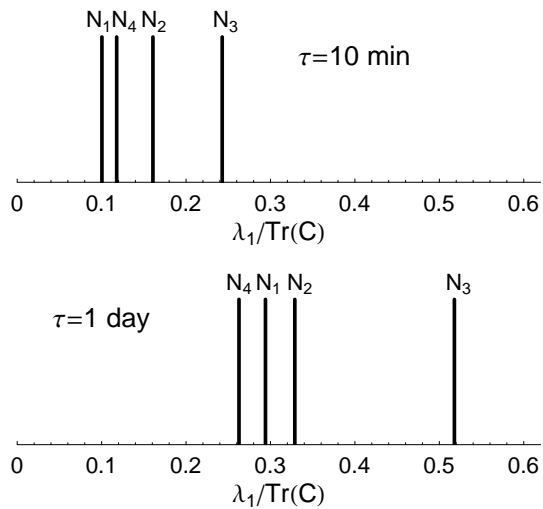}
\caption{Empirical eigenvalue spectrum of the correlation matrix ${\bf C}$
(vertical lines), calculated for 4 groups of companies over the period
17.11.2000-30.06.2005. Two time scale are presented: $\tau=10min$ (top)
and $\tau=360min$ (bottom).}
\end{figure}

By comparing the results in both panels of Figure 3, one can notice that
for the longer time scale the eigenvalues assume larger magnitudes than do
their counterparts for the shorter time scale. This can be a manifestation
of the Epps effect~\cite{epps79,kwapien04}, i.e. increase of market
cross-correlations with increasing time scale of the returns. In order to
verify this supposition, we systematically inspected the functional
dependence of $\lambda_1$ on time lag $\tau$ for a few distinct time
scales 10 min $< \tau <$ 900 min. Indeed, the results collected in Figure
4 confirm that the observed behaviour of the largest eigenvalue is a
consequence of the Epps effect. For all the groups, $\lambda_1$ increases
from small values for the shortest time lags to the group-specific
saturation levels for $\tau > 200$ min (compare with the American market
with similar saturation occurring for $\tau > 20$ min). It is noteworthy
that for Group 3 and Group 4 and for $\tau = 10$ min the largest
eigenvalue is comparable in size with the noise level ($\lambda_{\rm
max}$) predicted by RMT. This outcome resembles the analogous one obtained
for the American stock market~\cite{kwapien04}: the smaller is the
capitalization of a group of stocks, the less internally correlated and
more noisy is the group's evolution, what - in consequence with the Epps
effect - leads to a complete lack of actual inter-stock couplings for
sufficiently short time scales. This phenomenon is related to the fact
that investors need some time to fully react to new information and events
on a stock market. Stocks of smaller companies are traded less frequently
than stocks of large companies and therefore the amount of time needed to
develop couplings between such stocks is considerably larger.

\begin{figure}
\epsfxsize 12.5cm
\hspace{0cm}
\epsffile{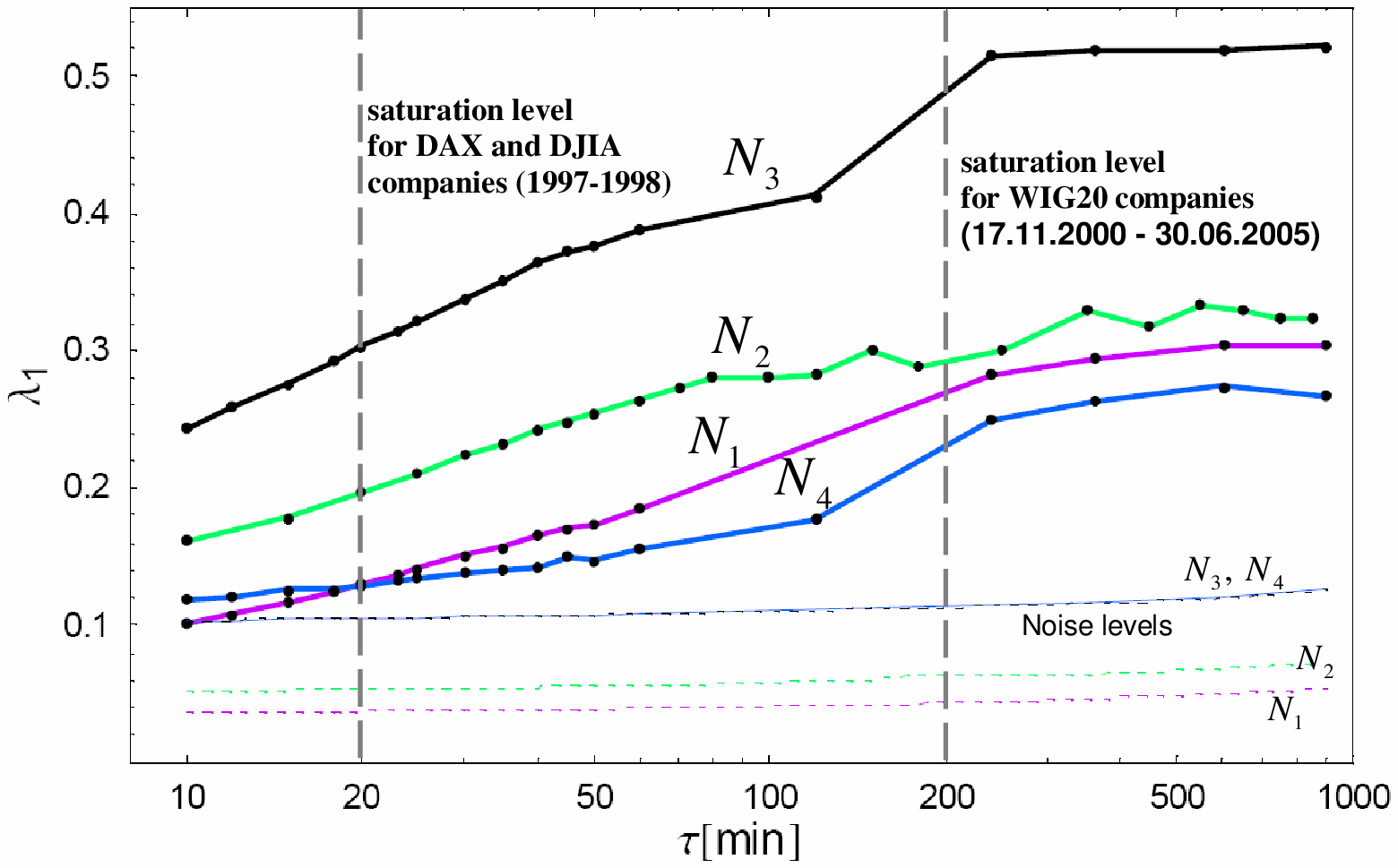}
\caption{Functional dependence of $\lambda_1$ on time lag $\tau$ for all
analyzed groups of stocks.}
\end{figure}

\section{Summary}

We investigated the inter-stock correlations for the relatively large
companies traded on Warsaw Stock Exchange and included in the WIG20 index.
We divided the full set of stocks into 4 groups depending on a particular
stock's capitalization and a time interval in which the stock was included
in WIG20. Our results show that the Polish stock market can basically be
expressed by an one-factor model with the fully developed couplings to
occur at time scales longer than half a trading day. Since these
properties are characteristic for small and emerging markets and since, on
the other hand, the Polish market reveals some features that are common to
well-developed markets ($q$-Gaussian structure of the returns
p.d.f.s~\cite{rak07}, multifractality~\cite{drozdz07a}), we arrive at
the conclusion that at present WSE is in a transition phase from being an
emerging market to becoming a fully-established one. Our analysis also
proved that the strength of correlations among stocks crucially depends on
their capitalization - this effect is universal for all the markets
investigated so far in literature.

\end{document}